\documentclass[twocolumn,nolinenumbers,trackchanges]{aastex7}

\begin{document}

\title{A search for chemical stratification and magnetic fields in eight field blue horizontal-branch stars\footnote{Based on observations obtained at the Canada-France-Hawaii Telescope (CFHT) which is operated by the National Research Council of Canada, the Institut National des Sciences de l'Univers of the Centre National de la Recherche Scientique of France, and the University of Hawaii. The observations at the Canada-France-Hawaii Telescope were performed with care and respect from the summit of Maunakea, which is a significant cultural and historic site.}} 

\author[orcid=0000-0000-0000-0001]{Issouf Kafando}
\affiliation{Royal Military College of Canada, 15 General Crerar Crescent, Kingston, ON, K7K 7B4, Canada}
\email[show]{issouf.kafando@rmc-cmr.ca}  

\author{Gregg Wade}
\affiliation{Royal Military College of Canada, 15 General Crerar Crescent, Kingston, ON, K7K 7B4, Canada}
\affiliation{Queen's University, Stirling Hall, 64 Bader Lane, Kingston, ON, K7L 3N6, Canada}
\email{gregg.wade@rmc.ca}

\author{Alexandre David-Uraz}
\affiliation{Department of Physics, Central Michigan University, Mt. Pleasant, MI 48859, USA}
\email{david7a@cmich.edu}

\author{Francis LeBlanc}
\affiliation{Université de Moncton, 18 avenue Antonine-Maillet, Moncton, N.-B, E1A 3E9, Canada}
\email{francis.leblanc@umoncton.ca}

\author{Carmelle Robert}
\affiliation{Université Laval $\&$ Centre de recherche en astrophysique du Qu\'ebec, 1045 avenue de la M\'edecine, Qu\'ebec, QC, G1V 0A6, Canada}
\email{carmelle.robert@phy.ulaval.ca}

\begin{abstract}

In this article, we present the results of an analysis of chemical abundances and a search for magnetic fields in eight field blue horizontal-branch (BHB) stars: BD\,+01$^{\circ}$0548, HD\,74721, HD\,86986, HD\,87112, HD\,93329, HD\,109995, HD\,161817, and HD\,167105. This study is based on high-resolution optical spectra obtained with ESPaDOnS at the Canada-France-Hawaii Telescope. We first calculated the average chemical element abundance, the rotational velocity, and the radial velocity of the stars using the ZEEMAN2 spectrum synthesis code with the PHOENIX model atmospheres. We then studied the abundances of titanium and iron inferred from individual lines in the spectra of each star and their variations with their predicted formation depths $\tau_{\rm 5000}$. Similarly to the BHB stars cooler than 11500~K observed in globular clusters, no clear observational evidence of vertical stratification is detected in the atmosphere of these stars. In the second part of this project,  we searched for the presence of a magnetic field in the stars applying the Least-Squares Deconvolution technique on the Stokes $I$ and $V$ spectra. The measured mean longitudinal magnetic field uncertainties, ranging from 8 to 30~G, effectively rule out the presence of an organized magnetic field in these targets with a strength larger than a few 100~G.

\end{abstract}


\section{Introduction}
The horizontal-branch stars are intermediate-mass stars that are burning helium in their cores \citep[e.g.][]{hoy55}. The distribution of stars on the horizontal branch primarily depends on the ratio between the core mass and the total mass of the star, as well as the mass loss occurring during the helium flash. When the ratio between the core mass and the total mass is high, the star is expected to end up on the left side of the horizontal branch, becoming a blue horizontal-branch (BHB) star. Otherwise, it will stabilize on the right side to become a red horizontal-branch (RHB) star. Between the BHB and RHB stars there are the RR Lyrae stars, which are variable stars \citep[e.g.][]{mae09}. Another group of objects, identified as extreme horizontal-branch (EHB) stars, move upward to the blue to become hot subdwarfs. It is expected that they will never reach the asymptotic giant branch (AGB) because they have undergone extreme mass loss. The EHB stars are often called ``AGB manqué'' \citep[e.g.][]{cruz96}. %\citep[]{mae09}. \par

Populations of horizontal-branch stars are observable in globular clusters. Also, they are present in populations of field stars, such as in the halo of the Galaxy. The field BHB stars are considered to be good distance indicators because of their high luminosities. They are ideal tracers for studying the profile of the outer stellar halo of the Milky Way \citep[]{2018MNRAS.481.5223T}.

Several studies have shown that the BHB stars hotter than approximately 11500 K observed in globular clusters exhibit abundance anomalies such as a helium underabundance and metal overabundances compared to the cluster's average abundance \citep{gla89, beh99, moe99}. 
Photometric jumps \citep{gru99} and gaps \citep[]{fer98} are also detected in the color-magnitude diagram of globular clusters near the critical value $T_\textrm{eff} \simeq$ 11500~K. A low rotation velocity \citep[$V \sin i \simeq$~10 km s$^{-1}$ or less ;][]{pet95, beh00} and a low spectroscopic gravity compared to stellar evolution models \citep[e.g.][]{cro88, moe95} are also observed anomalies for these BHB stars.\par

The low rotation velocity of hot BHB stars in globular clusters makes the stellar atmosphere hydrodynamically stable, conducive to the process of atomic diffusion \citep[]{mic70}, thus leading to abundance anomalies that result in the vertical stratification of certain chemical elements in their atmosphere. In the work of \citet{kha07} on the BHB star HD\,135485, nitrogen and sulfur show signatures of vertical abundance stratification in the stellar atmosphere. \citet{kha08, kha10} detected vertical stratification of iron in two BHB stars in the globular cluster NGC\,288, two stars in M\,13, and three stars in M\,15. \citet{kaf16} presented a detailed abundance analysis of five field BHB stars and detected vertical stratification of phosphorus in the star HD\,213781 ($T_\textrm{eff} =$ 13330 $\pm$ 350~K) classified by \citet{beh03} as a possible BHB star, and a marginal indication of vertical stratification of iron in Feige\,86 ($T_\textrm{eff} = 16110 \pm 450$ K). 

\citet{hui00} and \citet{leb09} developed atmosphere models that take into account the stratification caused by diffusion. These models suggest that stratification leads to a change in atmospheric structure, hence changes in colors, resulting in photometric anomalies. These models also helped explain the low spectroscopic gravities observed for each of the stars \citep[]{hui00, leb10}.

Because BHB stars show strong and narrow absorption lines in their spectra, and also because some of them possess chemical abundance anomalies similar to Ap/Bp stars \citep[e.g.][]{2007MNRAS.376..361F}, \citet{Elk98} measured the magnetic field and atmospheric parameters of eight BHB stars including our stars HD\,74721, HD\,86986, HD\,109995, and HD\,161817. \citet{pau19} used time-series photometric data, taken from ASAS, ASAS-SN, and SuperWASP surveys, to conduct a search for stellar spots on 30 field BHB stars, including our targets. 

This article presents an analysis of the spectral lines of eight field BHB stars to search for evidence of vertical stratification of elemental abundances and for Zeeman signatures in their atmosphere. The aim of this work is to increase the quantity of information on the physico-chemical parameters of the BHB stars. 

In Section 2, we present the sample of stars studied and the observational information. In Sections 3 and 4, we determine, respectively, the element abundance and the mean longitudinal magnetic field measurement for each star. In the last section, a discussion and a conclusion are followed.

\begin{deluxetable*}{lrcc}
%\tabletypesize{\scriptsize}
\tablewidth{0 pt} 
%\tablenum{1}
\tablecaption{Summary of ESPaDOnS observations. For each star studied, we list the observing date (Column~2), total exposure times (Column~3) and signal-to-noise ratio (S/N) per pixel (Column~4).  \label{tab:observational}}
\tablehead{
\colhead{Star} & \colhead{Date}& \colhead{ $t_{exp}$ (s)} & \colhead{S/N (pix$^{-1}$)} 
}
%\colnumbers
\startdata 
BD\,+01$^{\circ}$0548  &  13-08-2022   & 1310   & 266            \\
HD\,74721  &   21-01-2025  &  430 & 218        \\
HD\,86986  &   21-01-2025  &  235 & 125        \\
HD\,87112  &   21-01-2025  &  1060 & 168        \\
HD\,93329  &   21-01-2025  &  480 & 316        \\
HD\,109995  &   21-01-2025  &  250 & 437        \\
HD\,161817  &   13-08-2022  &  60 & 262        \\
HD\,161817  &   15-08-2022  &  60 & 311        \\
HD\,161817  &   16-08-2022  &  60 & 301       \\
HD\,167105 &   13-08-2022  &  960  & 426        \\\hline      
\enddata
\end{deluxetable*}

\begin{deluxetable*}{lrcccccc}
%%\tabletypesize{\scriptsize}
\tablewidth{0 pt} 
%%\tablenum{1}
\tablecaption{ Physical parameters of the field BHB stars. \label{tab:parameters}}.
\tablehead{
            &              &  & &This study   &   Behr(2003) & This study  &  Behr(2003)\\
\colhead{Star} & \colhead{$T_\textrm{eff}$}& \colhead{log $g$} & \colhead{[M/H]}& \colhead{$V \sin i$}& \colhead{$V \sin i$} &\colhead{$V_\textrm{r}$} & \colhead{$V_\textrm{r}$}\\
\colhead{} & \colhead{(K)} & \colhead{(cm s${^-}{^2}$)}& \colhead{} & \colhead{(km s$^{-1}$)}& \colhead{(km s$^{-1}$)}& \colhead{(km s$^{-1}$)}& \colhead{(km s$^{-1}$)}
}
%%\colnumbers
\startdata 
 BD\,+01$^{\circ}$0548  &   $8714_{-160}^{+235}$  &  $3.38_{-0.12}^{+0.20}$&$-2.00$  & $10.71\pm 1.25$ &  $ 10.2_{-0.8}^{+0.7}$  &  $-54.20 \pm 0.82$ &$-55.84 \pm 0.79$  \\
 HD\,74721  & $8810 \pm 110$  & $3.25 \pm 0.05$ & $-1.50$& $3.63 \pm 1.58$ &  $ 2.6_{-2.6}^{+1.4}$ & $+31.35 \pm 0.68$& $31.41 \pm 0.36$ \\
 HD\,86986  & $7980 \pm 30$  & $2.94 \pm 0.03$ & $-1.50$& $8.86 \pm 2.01 $ & $ 9.2_{-0.8}^{+0.9}$  & $+9.91 \pm 1.16$&$10.13 \pm 0.50$  \\
 HD\,87112  & $9440 \pm 25$  & $3.39 \pm 0.02$ & $-1.50$& $7.43 \pm 1.35 $ & $ 7.2_{-1.6}^{+1.9}$  & $-173.05 \pm 0.68$& $-172.55 \pm 0.69$ \\
 HD\,93329  & $8180 \pm 40$  & $2.95 \pm 0.02$ & $-1.50$& $9.55 \pm 1.40$ & $ 9.6_{-1.3}^{+1.3}$  & $+204.84 \pm 0.86$&$+204.97 \pm 0.71$  \\
 HD\,109995  & $8460 \pm 40$  & $3.17 \pm 0.02$ & $-1.50$& $23.25 \pm 1.63$ & $ 22.9_{-1.7}^{+2.0}$  & $-130.10 \pm 0.96$& $-126.31 \pm 1.87$  \\
 HD\,161817  &   $7400 \pm 35$  &  $2.84 \pm 0.07$& $-1.50$  & $15.80\pm 1.68$ & $ 15.2_{-0.7}^{+0.8}$  &$-363.61\pm 1.14$ &$-362.65 \pm 1.29$   \\
 HD\,167105  &   $8790 \pm 70$  &  $3.24 \pm 0.03$ &$-1.50$ & $20.34\pm 1.41$ & $20.0_{-1.8}^{+1.9}$ &$ -174.17 \pm 1.09$  &$-171.82 \pm 1.36$   \\\hline  
 \enddata
 \tablecomments{Effective temperature and surface gravity of BD\,+01$^{\circ}$0548 are from \citet{beh03}, and of the other stars from \citet{kaf17}. The [M/H] values are from \citet{beh03}}
 \end{deluxetable*}

\begin{deluxetable*}{lcccc}
%\digitalasset
%\tabletypesize{\scriptsize}
\tablewidth{0 pt} 
%\tablenum{1}
\tablecaption{List of spectral lines used for each star studied. \label{tab:finallist}}
\tablehead{
\colhead{$\lambda$ (\AA)  }&\colhead{log($\emph{gf}$)} & \colhead{log $(N_{\rm ion}/N_{\rm tot}$)}&  \colhead{log$(\tau_{\rm 5000})$} &\colhead{$E_{\rm low}$(cm$^{-1}$)}
}
\startdata
 {}    & {}  & (a) BD\,+01$^{\circ}$0548  &  {}   & {} \\\hline
  {}   & {}  &  N\,{\sc i}  & {}  &   {}        \\
8629.2350 & $+0.077$ &  $-5.2761 \pm 0.0317$  &  $-2.9748$  &  86220.51  \\
8680.2830  &  $+0.359$ & $-5.5216 \pm 0.0161$ & $-3.4236$ & 83364.62         \\  
8683.4030  &  $+0.105$ & $-5.5781 \pm 0.0284$ & $-3.1987$ & 83317.83         \\
\nodata  &  \nodata & \nodata & \nodata & \nodata         \\\hline
\enddata
\tablecomments{Table~$\ref{tab:finallist}$ is published in its entirety in the electronic edition of the {\it Astrophysical Journal}.  A portion is shown here 
for guidance regarding its form and content.}
\end{deluxetable*}

\section{STARS STUDIED AND OBSERVATIONS} \label{sec:style}

We selected the field BHB stars BD\,+01$^{\circ}$0548, HD\,74721, HD\,86986, HD\,87112, HD\,93329, HD\,109995, HD\,161817, and HD\,167105. These stars have already been studied by \citet{ade86}, \citet{ade87}, \citet{ade94},  \citet{ade96}, \citet{kin00}, \citet{beh03}, and \citet{kaf17} to determine chemical abundances, effective temperature, surface gravity, and radial and rotational velocities. 

We obtained spectra with the echelle spectrograph ESPaDOnS (see Table~$\ref{tab:observational}$ for the summary of observations) at the Canada-France-Hawaii Telescope (CFHT) in August 2022 and January 2025. ESPaDOnS offers the complete spectrum in the visible region, covering the wavelength range from 3700 to 10500 \AA, with a spectral resolution R $\approx$ 65000 in spectropolarimetric mode. Each polarimetric sequence consisted of four individual subexposures taken in different polarimeter configurations. From each set of four subexposures, we derived a mean Stokes $\mathrm{V}$ spectrum that was used to estimate the longitudinal magnetic field. Also, from each set of four subexposures, we computed a single combined spectrum to estimate the elemental abundances of the stars studied. CFHT spectra were reduced and normalized with the Upena pipeline, which uses the routines from Libre-ESpRIT \citep[]{don97}. A second normalization of the observed spectrum was carried out by fitting a polynomial of degree 5 through selected points for each of the 40 orders in the echelle. 

We report in  Table~$\ref{tab:parameters}$ the effective temperature ($T_\textrm{eff}$) and surface gravity (log $g$) of our target sample. For BD\,+01$^{\circ}$0548 we used the values of \citet{beh03} while we used those determined by \citet{kaf17} for the other stars. The metallicity [M/H] given for each star in Table~$\ref{tab:parameters}$ corresponds to the iron abundance calculated by \citet{beh03}. Measurements of the rotational velocity ($V \sin i$) and radial velocity ($V_\textrm{r}$) are from this work (see Section 3.1). Values from \citet{beh03} are also given in Table~$\ref{tab:parameters}$ for comparison. 

The stars studied here are cooler than 11500~K which is the effective temperature threshold above which the detection of vertical stratification of iron is observed in the BHB stars of globular clusters \citep[]{kha10}.

\begin{deluxetable*}{lcccc}
%\tabletypesize{\scriptsize}
\tablewidth{700 pt} 
%\tablenum{1}
\tablecaption{Average chemical abundances log $(N_{\rm ion}/N_{\rm tot}$) of BD\,+01$^{\circ}$0548. \label{tab:abbd010548}}
\tablehead{
%& &    & \colhead{Adelman et al.} &\colhead{Kinman et al.}\\ 
\colhead{Ion}  & \colhead{\textit{N}}  & \colhead{This study}
%  &   &\colhead{log $(N_{\rm ion}/N_{\rm tot}$)} &\colhead{log $(N_{\rm ion}/N_{\rm tot}$)}
}
%\colnumbers
\startdata 
%BD\,+01$^{\circ}$0548 &           &              & &  &\colhead{Adelman et al. (1986,1987,1994)}  \\ \hline
%\colhead{Ion} & \colhead{N}& \colhead{This study} & \hline
 N\,{\sc i}  & 3 & $-5.46 \pm 0.16 $ \\
 O\,{\sc i}  & 3 & $-3.70 \pm 0.45$    \\
 Mg\,{\sc \i}  & 2 & $-6.24 \pm 0.04$   \\
 Mg\,{\sc ii}  & 1 & $-6.20$    \\
 Si\,{\sc ii}  &  7 & $-6.16 \pm 0.21$  \\
 Ca\,{\sc ii}  & 4 & $-7.71 \pm 0.28$ &  \\
 Ti\,{\sc ii}  & 13 &$-8.84 \pm 0.15$ &  \\
 Fe\,{\sc i}  &  3  & $-6.78 \pm 0.13$  &  \\
 Fe\,{\sc ii} & 13 & $-6.55 \pm 0.23$ &  \\
%\hline
\enddata
\end{deluxetable*}

%%%%%%%%%%%%%%%%%%%%%%%%%%%%%%%%%%%%%%%%%%%%%%%%%
\begin{deluxetable*}{ccccc}
%\tabletypesize{\scriptsize}
\tablewidth{0 pt} 
%\tablenum{1}
\tablecaption{Average chemical abundances log $(N_{\rm ion}/N_{\rm tot}$) of HD\,74721. \label{tab:abhd74721}}
\tablehead{
\colhead{Ion}& \colhead{\textit{N}}& \colhead{This study} & \colhead{Adelman et al.} &\colhead{Kinman et al.}\\ 
&   &  & \colhead{et al. (1996)} & \colhead{et al. (2000)}
%  &   &\colhead{log $(N_{\rm ion}/N_{\rm tot}$)} & \colhead{log $(N_{\rm ion}/N_{\rm tot}$)} & \colhead{log $(N_{\rm ion}/N_{\rm tot}$)}
}
%\colnumbers
\startdata 
 C\,{\sc i}   &  4 & $-5.12 \pm 0.12$ &  \nodata &  \nodata\\
 N\,{\sc i}   &  9 & $-4.83 \pm 0.10$  &  \nodata &  \nodata \\
 O\,{\sc i}   &  5 & $-3.88 \pm 0.17$ &  \nodata &  \nodata \\
 Na\,{\sc i}  &  3 & $-7.19 \pm 0.06$ &  \nodata &  \nodata \\
 Mg\,{\sc i}  &  6 & $-5.33 \pm 0.55$ &  $-5.47$ &  \nodata  \\
 Mg\,{\sc ii} &  6 & $-5.38 \pm 0.10$ &   $-5.37 \pm 0.22$ & $-5.64$ \\
 Al\,{\sc i}  &  1 & $-7.23$ &   $-7.33 \pm 0.18$ &  \nodata \\
 Si\,{\sc ii} &  9 & $-5.31 \pm 0.28$ &   $-5.58 \pm 0.06$ &  \nodata\\
 S\,{\sc i}   &  2 & $-5.31 \pm 0.24$ &   \nodata &  \nodata\\
 Ca\,{\sc i}  &  5 & $-7.28 \pm 0.07$ &   $-7.08 \pm 0.27$ &  \nodata  \\
 Ca\,{\sc ii} &  3 & $-7.15 \pm 0.17$ &   $-6.90$ &  \nodata\\
 Sc\,{\sc ii} &  8 & $-10.41 \pm 0.11$&  $-10.35 \pm 0.16$ & $-10.08$\\
 Ti\,{\sc ii} & 42 & $-8.05 \pm 0.33$ &  $-8.02 \pm 0.21$ & $-8.11 \pm 0.13$ \\
 Cr\,{\sc ii} &  9 & $-7.77 \pm 0.10$ &   $-7.78 \pm 0.25$ & $-7.64$\\
 Fe\,{\sc i}  & 53 & $-6.13 \pm 0.21$ &   $-5.99 \pm 0.23$ & $-5.95 \pm 0.10$\\
 Fe\,{\sc ii} & 36 & $-5.83 \pm 0.23$ &   $-5.90 \pm 0.24$ & $-5.97 \pm 0.08$\\
 Sr\,{\sc ii} & 2  & $-10.75 \pm 0.08$ &  $-10.82$ &  \nodata\\
 Zr\,{\sc ii} & 1 & $-10.45$ &  $-10.43$ &  \nodata\\
 Ba\,{\sc ii} &  2 & $-11.53 \pm 0.09$ & $-11.58$ &  \nodata\\
%\hline
\enddata
\end{deluxetable*}
%%%%%%%%%%%%%%%%%%%%%%%%%%%%%%%%%%%%
\begin{deluxetable*}{lcccc}
%\tabletypesize{\scriptsize}
\tablewidth{700 pt} 
%\tablenum{1}
\tablecaption{Average chemical abundances log $(N_{\rm ion}/N_{\rm tot}$) of HD\,86986. \label{tab:abhd86986}}
\tablehead{
\colhead{Ion} & \colhead{\textit{N}}& \colhead{This study}   & \colhead{Adelman et al.} &\colhead{Kinman et al.}\\ 
  &   &  & \colhead{(1996)} & \colhead{(2000)}
%   &   &\colhead{log $(N_{\rm ion}/N_{\rm tot}$)} & \colhead{log $(N_{\rm ion}/N_{\rm tot}$)} & \colhead{log $(N_{\rm ion}/N_{\rm tot}$)}
}
%\colnumbers
\startdata 
C\,{\sc i}   &5 & $-5.23 \pm 0.11$  &   \nodata &  \nodata \\
N\,{\sc i}   &3 & $-5.28 \pm 0.21$ &    \nodata &  \nodata \\
O\,{\sc i}   &5 & $-3.47 \pm 0.96$ &    \nodata &  \nodata \\
Mg\,{\sc i}  &6 & $-5.50 \pm 0.53$ &    $-5.81$  &  \nodata \\
Mg\,{\sc ii} &  2 & $-5.66 \pm 0.08$&   $-5.54 \pm 0.37$ & $-5.72$ \\
Al\,{\sc i}  &  1 & $-7.44$ &   $-7.96 \pm 0.22$ &  \nodata \\
Si\,{\sc ii} &  5 & $-5.96 \pm 0.28$ &   $-5.89 \pm 0.20$ &  \nodata\\
S\,{\sc i}   &  1 & $-5.57$ &   \nodata &  \nodata\\
Ca\,{\sc i}  &  5 & $-6.87 \pm 0.38$ &   $-7.33 \pm 0.11$  & $-7.07$ \\
Ca\,{\sc ii} &  5 & $-6.98 \pm 0.38$ &   $-7.15$ &  \nodata\\
Sc\,{\sc ii} &  4 & $-10.33 \pm 0.22$ & $-10.61 \pm 0.13$ &  \nodata\\
Ti\,{\sc ii} & 35 & $-8.02 \pm 0.55$  & $-8.37 \pm 0.19$ & $-8.32 \pm 0.17$\\
Cr\,{\sc i}  &  3 & $-7.97 \pm 0.13$ &   $-8.37 \pm 0.19$ &  \nodata \\
Cr\,{\sc ii} &  5 & $-7.98 \pm 0.13$ &   $-7.84 \pm 0.26$ & $-7.79$\\
Fe\,{\sc i}  & 44 & $-6.12 \pm 0.21$ &   $-6.44 \pm 0.14$ & $-6.36 \pm 0.10$\\
Fe\,{\sc ii} & 23 & $-6.06 \pm 0.20$ &  $-6.24 \pm 0.13$ & $-6.34 \pm 0.10$\\
Sr\,{\sc ii} &  2 & $-10.30 \pm 0.42$ & $-11.44$ &  \nodata \\
Ba\,{\sc ii} &  2 & $-11.69 \pm 0.07$ & $-12.06$ & $-11.85$\\
%\hline
\enddata
\end{deluxetable*}
%%%%%%%%%%%%%%%%%%%%%%%%%%%%%%%%%%%
\begin{deluxetable*}{lcccc}
%\tabletypesize{\scriptsize}
\tablewidth{700 pt} 
%\tablenum{1}
\tablecaption{Average chemical abundances log~$(N_{\rm ion}/N_{\rm tot}$) of HD\,87112. \label{tab:abhd87112}}
\tablehead{
\colhead{Ion}& \colhead{\textit{N}}  &  \colhead{This study} &\colhead{Kinman et al.}\\ 
&   &   & \colhead{(2000)}
% &  &\colhead{log $(N_{\rm ion}/N_{\rm tot}$)} & \colhead{log $(N_{\rm ion}/N_{\rm tot}$)} & \colhead{log $(N_{\rm ion}/N_{\rm tot}$)}
}
%\colnumbers
\startdata 
C\,{\sc i}   &  1 & $-4.67$  &  \nodata \\
N\,{\sc i}   &  5 & $-4.96 \pm 0.18$   &  \nodata \\
O\,{\sc i}   &  6 & $-3.99 \pm 0.22$  & \nodata  \\
Na\,{\sc i}  &  2 & $-7.24 \pm 0.11$  & \nodata  \\
Mg\,{\sc i}  &  6 &$-5.65 \pm 0.12$  &  \nodata  \\
Mg\,{\sc ii} &  6 &$-5.44 \pm 0.08$  &  $-5.55$ \\
Si\,{\sc ii} &  10 & $-5.58 \pm 0.22$  &  \nodata \\
Ca\,{\sc i}  &  1 & $-7.58$  &  \nodata   \\
Ca\,{\sc ii} &  2 & $-7.56 \pm 0.50$   &  \nodata \\
Sc\,{\sc ii} &  1 & $-10.99$  &  \nodata \\
Ti\,{\sc ii} & 15 & $-8.47 \pm 0.12$  &  $-8.19 \pm 0.11$ \\
Cr\,{\sc ii} &  4 & $-8.04 \pm 0.11$  &  $-7.62$ \\
Fe\,{\sc i}  &  8 & $-6.23 \pm 0.07$  &  $-5.92$ \\
Fe\,{\sc ii} & 23 & $-6.00 \pm 0.29$  &  $-6.08 \pm 0.11$ \\
%\hline
\enddata
\end{deluxetable*}
%%%%%%%%%%%%%%%%%%%%%%%%%%%
\begin{deluxetable*}{lccccc}
%\tabletypesize{\scriptsize}
\tablewidth{700 pt} 
%\tablenum{1}
\tablecaption{Average chemical abundances log $(N_{\rm ion}/N_{\rm tot}$) of HD\,93329. \label{tab:abhd93329}}
\tablehead{
\colhead{Ion}  &\colhead{\textit{N}}  &  \colhead{This study}  & \colhead{Adelman et al.} &\colhead{Kinman et al.}\\ 
  &   &  & \colhead{(1996)} & \colhead{(2000)}
% &  &  &\colhead{log $(N_{\rm ion}/N_{\rm tot}$)} & \colhead{log $(N_{\rm ion}/N_{\rm tot}$)} & \colhead{log $(N_{\rm ion}/N_{\rm tot}$)}
}
%\colnumbers
\startdata 
C\,{\sc i}   &  4 & $-5.13 \pm 0.32$ &   \nodata &  \nodata \\
N\,{\sc i}   &  6 & $-5.03 \pm 0.10$ &   \nodata &  \nodata\\
O\,{\sc i}   &  3 & $-4.04 \pm 0.04$ &   \nodata &  \nodata \\
Na\,{\sc i}  &  2 & $-6.98 \pm 0.32$&    \nodata &  \nodata \\
Mg\,{\sc i}  &  5 & $-5.57 \pm 0.10$ &   $-5.51$ &  \nodata  \\
Mg\,{\sc ii} &  3 & $-5.38 \pm 0.18$ &  $-5.33 \pm 0.16$ & $-5.27$\\
Al\,{\sc i}  &  1 & $-7.33$ &  $-7.56 \pm 0.18$ &  \nodata  \\
Si\,{\sc ii} &  7 & $-5.49 \pm 0.31$ &   $-5.75$ &  \nodata \\
S\,{\sc i}   &  2 & $-4.81 \pm 0.99$ &   \nodata &  \nodata\\
Ca\,{\sc i}  &  11 &$-6.89 \pm 0.30$ &   $-7.06 \pm 0.20$ &  \nodata  \\
Ca\,{\sc ii} &  7 & $-6.78 \pm 0.44$ &  $-6.81$ &  \nodata \\
Sc\,{\sc ii} &  8 & $-10.11 \pm 0.34$ & $-10.21 \pm 0.14$ & $-10.25 \pm 0.03$\\
Ti\,{\sc ii} & 46 & $-7.79 \pm 0.65$ &  $-8.00 \pm 0.20$  & $-7.84 \pm 0.19$\\
V\,{\sc ii}  &  3 & $-9.28 \pm 0.09$ &  $-9.28 \pm 0.08$ &  \nodata\\
Cr\,{\sc i}  &  4 &$-8.00 \pm 0.04$ &  $-7.97 \pm 0.07$ &  \nodata\\
Cr\,{\sc ii} &  15 & $-7.78 \pm 0.11$ & $-7.58 \pm 0.15$ & $-7.51$\\
Fe\,{\sc i}  & 77 & $-5.95 \pm 0.28$ & $-5.95 \pm 0.25$ & $-5.86 \pm 0.08$\\
Fe\,{\sc ii} & 41 & $-5.77 \pm 0.35$ &  $-5.93 \pm 0.13$ & $-5.87 \pm 0.06$\\
Ni\,{\sc i}  &  2 & $-7.24 \pm 0.09$ &  \nodata &  \nodata\\
Sr\,{\sc ii} &  2 & $-9.87 \pm 0.39$ &  $-11.33$ &  \nodata\\
Y\,{\sc ii}  &  2 & $-11.24 \pm 0.13$ & $-10.86$ &  \nodata\\
Zr\,{\sc ii} &  2 & $-10.40 \pm 0.19$ & $-10.42 \pm 0.11$ &  \nodata\\
Ba\,{\sc ii} &  2 & $-11.38 \pm 0.03$ &  $-11.56$ & $-11.27$\\
%\hline
\enddata
\end{deluxetable*}
%%%%%%%%%%%%%%%%%%%%%%%%%%%

\begin{deluxetable*}{lccccc}
%\tabletypesize{\scriptsize}
\tablewidth{700 pt} 
%\tablenum{1}
\tablecaption{Average chemical abundances log $(N_{\rm ion}/N_{\rm tot}$) of HD\,109995. \label{tab:abhd109995}}
\tablehead{
  \colhead{Ion}&\colhead{\textit{N}} &  \colhead{This study}  & \colhead{Adelman et al.} &\colhead{Kinman et al.}\\ 
    &   &  & \colhead{(1986, 1987)} & \colhead{(2000)}
% &   &\colhead{log $(N_{\rm ion}/N_{\rm tot}$)} & \colhead{log $(N_{\rm ion}/N_{\rm tot}$)} & \colhead{log $(N_{\rm ion}/N_{\rm tot}$)}
}
%\colnumbers
\startdata 
N\,{\sc i}   &  3 & $-5.02 \pm 0.24$ &  $-4.38$ &  \nodata \\
O\,{\sc i}   &  3 & $-3.33 \pm 1.16$ &  $-2.44$ &  \nodata \\
Na\,{\sc i}  &  2 &  $-7.14 \pm 0.01$  &  \nodata &  \nodata \\
Mg\,{\sc i}  &  7 & $-5.19 \pm 0.54$ &  $-5.84$  &  \nodata \\
Mg\,{\sc ii} &  2 & $-5.77 \pm 0.26$ &  $-5.40$  & $-5.84$\\
Al\,{\sc i}  &  1 & $-7.31$ &  $-7.93$ &  \nodata \\
Si\,{\sc ii} &  6 & $-5.94 \pm 0.30$ &  $-5.69$ &  \nodata \\
Ca\,{\sc i}  &  3 & $-6.80 \pm 0.10$ & $-7.50$ &  \nodata  \\
Ca\,{\sc ii} &  3 & $-7.02 \pm 0.13$ &  $-7.67$ &  \nodata\\
Sc\,{\sc ii} & 1 &  $-10.22$ &  $-10.78$ &  \nodata\\
Ti\,{\sc ii} & 30 & $-7.98 \pm 0.44$ &  $-8.22$ & $-8.31 \pm 0.12$\\
Cr\,{\sc i}  &  2 & $-7.77 \pm 0.16$ &  $-8.35$ &  \nodata\\
Cr\,{\sc ii} &  4 & $-7.84 \pm 0.17$ &  $-7.44$ &  \nodata \\
Fe\,{\sc i}  & 17 & $-5.81 \pm 0.27$ &  $-6.46$ & $-6.24 \pm 0.12$ \\
Fe\,{\sc ii} & 24 & $-5.96 \pm 0.49$ &  $-6.18$ & $-6.28 \pm 0.13$\\
Sr\,{\sc ii} &  2 & $-10.95 \pm 0.12$ &  $-11.57$ &  \nodata\\
%\hline
\enddata
\end{deluxetable*}
%%%%%%%%%%%%%%%%%%%%%%%%%%%
\begin{deluxetable*}{lccccc}
%\tabletypesize{\scriptsize}
\tablewidth{700 pt} 
%\tablenum{1}
\tablecaption{Average chemical abundances log $(N_{\rm ion}/N_{\rm tot}$) of HD\,161817. \label{tab:abhd161817}}
\tablehead{
\colhead{Ion} &\colhead{\textit{N}} &  \colhead{This study}  & \colhead{Adelman et al.} &\colhead{Kinman et al.}\\ 
    &   &  & \colhead{(1986, 1987)} & \colhead{(2000)}
%  &   &\colhead{log $(N_{\rm ion}/N_{\rm tot}$)} & \colhead{log $(N_{\rm ion}/N_{\rm tot}$)} & \colhead{log $(N_{\rm ion}/N_{\rm tot}$)}
}
%\colnumbers
\startdata 
C\,{\sc i}   &  4 & $-4.79 \pm 0.71$ &  $-4.60$ &  \nodata\\
N\,{\sc i}   &  5 & $-5.03 \pm 0.08$  &  $-4.45$ &  \nodata \\
O\,{\sc i}   &  2 & $-2.39 \pm 0.41$ &  $-2.62$ &  \nodata \\
Na\,{\sc i}  &  1 & $-5.64$  &  \nodata &  \nodata \\
Mg\,{\sc i}  &  4 & $-5.45 \pm 0.24$ &  $-5.57$  &  \nodata \\
Mg\,{\sc ii} &  2 & $-5.52 \pm 0.28$ &  $-5.02$ & $-5.69$\\
Al\,{\sc i}  &  2 & $-7.42 \pm 0.03$ &  $-7.62$  &  \nodata\\
Si\,{\sc ii} &  7 & $-5.55 \pm 0.26$ &  $-5.66$ &  \nodata\\
Ca\,{\sc i}  &  11 & $-6.67 \pm 0.40$ & $-7.01$ & $-6.99$  \\
Ca\,{\sc ii} &  3 & $-6.52 \pm 0.52$ &  $-7.01$ &  \nodata\\
Ti\,{\sc ii} & 44 & $-7.52 \pm 0.76$ &  $-8.11$ & $-8.13 \pm 0.24$\\
Cr\,{\sc ii} &  9 & $-7.74 \pm 0.12$ &  $-7.39$ &  \nodata\\
Fe\,{\sc i}  & 68 & $-5.51 \pm 0.64$ &  $-6.07$ & $-6.06 \pm 0.12$\\
Fe\,{\sc ii} & 30 & $-5.54 \pm 0.48$ &  $-6.00$ & $-6.12 \pm 0.07 $ \\
Ni\,{\sc i}  &  1 & $-7.12$ &  $-7.25$ & \nodata \\
Ba\,{\sc ii} &  2 & $-10.91 \pm 0.22$ & $-11.70$ & $-11.60$ \\
% & & & & &\\
%\hline
\enddata
\end{deluxetable*}
%%%%%%%%%%%%%%%%%%%%%%%%%%%
\begin{deluxetable*}{lccccc}
%\tabletypesize{\scriptsize}
\tablewidth{700 pt} 
%\tablenum{1}
\tablecaption{Average chemical abundances log $(N_{\rm ion}/N_{\rm tot}$) of HD\,167105.  \label{tab:abhd167105}}
\tablehead{
\colhead{Ion} & \colhead{\textit{N}} &\colhead{This study} & \colhead{Adelman et al.} &\colhead{Kinman et al.}\\ 
    &   &  & \colhead{(1994)} & \colhead{(2000)}
%  &  &   &\colhead{log $(N_{\rm ion}/N_{\rm tot}$)} & \colhead{log $(N_{\rm ion}/N_{\rm tot}$)} & \colhead{log $(N_{\rm ion}/N_{\rm tot}$)}
}
%\colnumbers
\startdata 
N\,{\sc i}   & 4 & $-4.75 \pm 0.16$ & \nodata   &  \nodata       \\
O\,{\sc i}   & 3 &$-2.61 \pm 0.79$ &  \nodata   &  \nodata       \\
Mg\,{\sc i}  & 4 & $-5.39 \pm 0.41$ &   \nodata &\nodata  \\
Mg\,{\sc ii} & 1 & $-5.50$ &  \nodata & $-5.77$   \\
Si\,{\sc ii} & 7 & $-5.58 \pm 0.21$ &  $-5.86 \pm 0.01$  &  \nodata       \\
Ca\,{\sc ii} & 2 & $-7.25 \pm 0.27$ &  $-7.42$   &  \nodata     \\
Sc\,{\sc ii} & 1 & $-10.63$ & $-11.00$   &  \nodata   \\
Ti\,{\sc ii} & 15 & $-7.88 \pm 0.28$ & $-8.39 \pm 0.07$ & $-8.20 \pm 0.14$ \\
Cr\,{\sc ii} & 3 &  $-7.77 \pm 0.08$ &  $-7.80 \pm 0.13$  & $-7.80$       \\
Fe\,{\sc i}  & 7 &  $-5.82 \pm 0.44$ &  $-6.40 \pm 0.19$ & $-6.08 \pm 0.05$ \\
Fe\,{\sc ii} & 18 & $-5.77 \pm 0.48$ & $-6.18 \pm 0.15$ & $-6.12 \pm 0.12$ \\
%\hline
\enddata
\end{deluxetable*}
%%%%%%%%%%%%%%%%%%%%%%%%%%%

\begin{figure*}
\centering
%\plotone{fitTi2.png}
\includegraphics[scale=0.75]{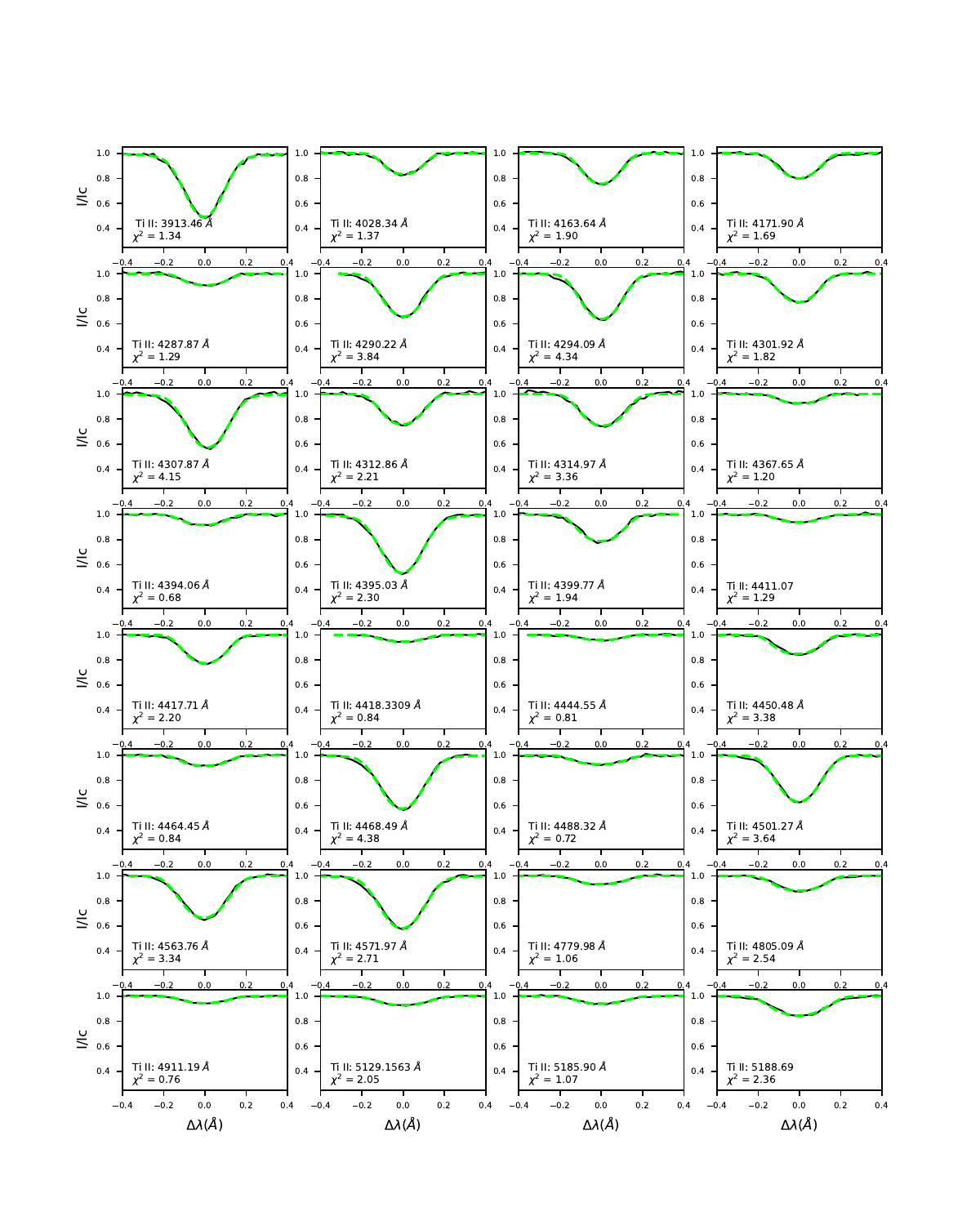}
\caption{Examples of ZEEMAN2 profile fitting results for 32 Ti II lines in HD\,93329. The observed profiles (black curves) are compared to the simulated ones (lime dotted curves).}
\label{fig:fitti2}
\end{figure*}
                                                                                                    
\section{ABUNDANCE ANALYSIS} \label{sec:cite}
\subsection{Procedure for the spectral analysis} \label{sec:procedure}

To identify the atomic lines present in the observed spectra we relied on the Vienna Atomic Line Database (VALD) of \citet{kup99}. The  ZEEMAN2 code \citep[]{lan88,wa01,kha06} was used to simulate each chosen line profile individually. The code allows for an automatic minimisation of the model parameters using the $\textit{downhill simplex method}$ \citep[]{pre92}. To analyze the vertical abundance stratification we considered the method which allows to determine the abundance of an element from an independent analysis of each line profile relative to the optical depth calculated at $5000$\,\AA \,\,($\tau_{5000}$) for the selected line profiles. The model atmosphere used contains 50 layers, and for every layer we calculate the line optical depth $\tau_{l}$ at the line centre using the following equation:
\begin{equation}
    \tau_{l}(m) = \tau_{l}(m-1) + \{1.0 +\kappa(m)\}\times\{\tau_{wl}(m) - \tau_{wl}(m-1)\}, 
\end{equation}
where $m > 1$ is the atmospheric level, $\kappa(m)$ is the ratio between the opacity at line centre and the continuum opacity, and $\tau_{wl}(m)$ is the optical depth calculated at a selected wavelength. We assume that the profile is formed mainly at line optical depth $\tau_{l} = 1$. This depth corresponds to a continuum optical depth $\tau_{5000}$ of a certain value that in turn corresponds to a given layer of the stellar atmosphere model \citep[]{kha07,kha10}.  

Each line profile analyzed has been cut (normally no wider than 2\,\AA) and normalized individually prior to modelling. The model atmospheres used in our study were calculated with the PHOENIX atmosphere code \citep[]{hau99} in the local thermodynamic equilibrium (LTE) mode. The effective temperature, surface gravity, and metallicity used for each model were adapted for each star as given in Table~$\ref{tab:parameters}$. We followed the same procedure as done for the spectral analysis by \citet{kaf16}. Only unblended lines and lines blended with the same species are treated in this study. 

To determine the abundance of a chemical element, we simultaneously fit the available line profile using the abundance of the element, the radial velocity, and the rotational velocity (see Fig.~$\ref{fig:fitti2}$ for examples of fits for HD\,93329). As \citet{kaf16}, taking into account other sources of uncertainties related to the model effects and the atomic data, we multiply the abundance uncertainties calculated by ZEEMAN2 by a factor 10.  In Table~$\ref{tab:parameters}$, we also report the radial and rotational velocities obtained here and those found by \citet{beh03} for the eight stars studied. In comparison, our values of both parameters generally agree well within the uncertainties with those obtained by \citet{beh03}.

%Our results are in agreement with the velocities computed by \citet{beh03}. 

Table~$\ref{tab:finallist}$  presents the final list of the atomic lines analysed for each star studied here, along with their wavelength $\lambda$, oscillator strength log$gf$, lower energy level $E_\textrm{low}$, and values obtained for their abundance and their line formation depth (the complete version of this table is found exclusively in the online version of the paper).

\begin{figure*}[ht!]
\centering
\includegraphics[scale=0.50]{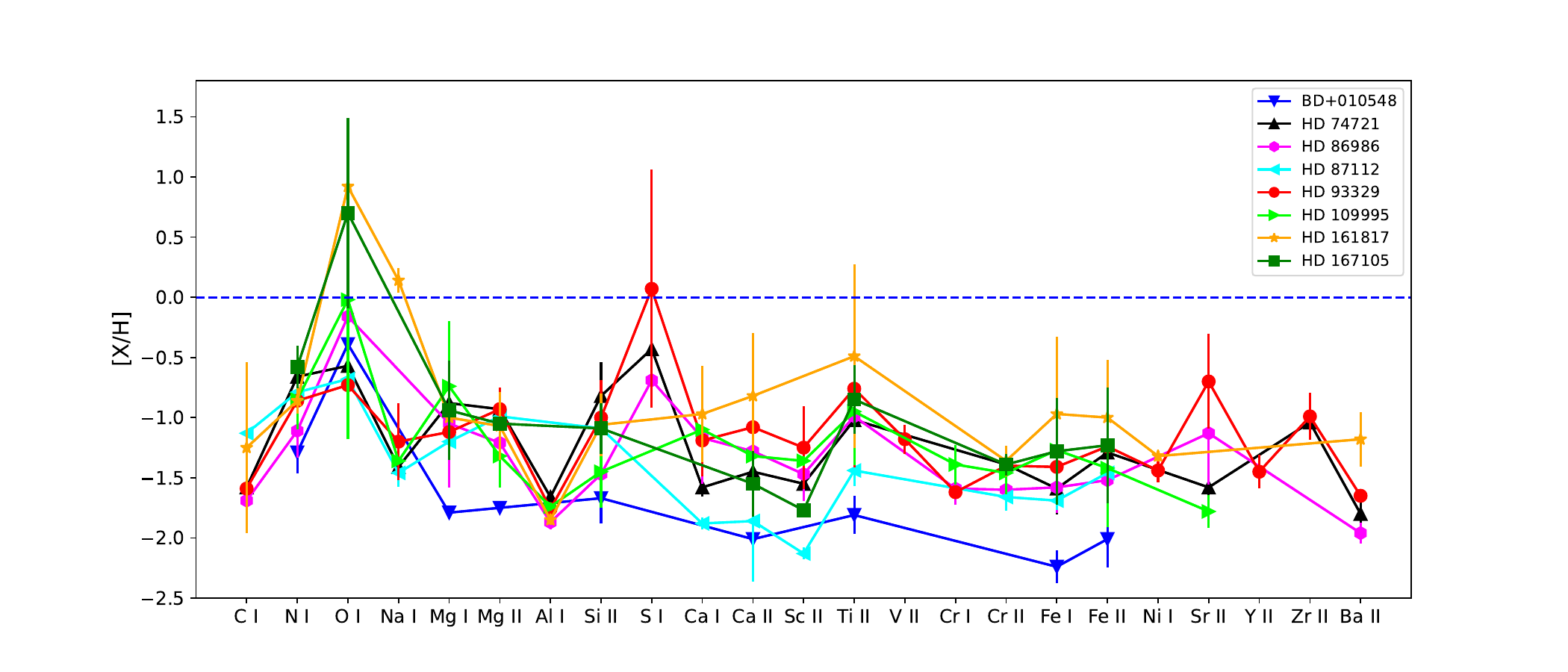}
\caption{The average abundances obtained relative to the solar abundance for the eight field BHB stars studied in this paper.}
\label{fig:anomalies}
\end{figure*}

\subsection{Average abundances}

 In comparison, we report in Tables~$\ref{tab:abbd010548}$ to $\ref{tab:abhd167105}$ our results for the average chemical abundance, log $(N_{\rm ion}/N_{\rm tot}$), and those obtained in previous studies. Figure~$\ref{fig:anomalies}$ shows the average abundance [X/H] obtained in our work relative to the solar abundance \citep[taken from][]{asp21} as a function of the element for the stars studied.

 For BD\,+01$^{\circ}$0548, Table~$\ref{tab:abbd010548}$ gives the average abundance of nine ions. However, for comparison, only the abundance of magnesium and iron computed by \citet{beh03} are available. \citet{beh03} used high spectral resolution data from the echelle spectrograph at McDonald Observatory and HIRES at Keck~I. For the analysis of the stellar spectra, \citet{beh03} used the spectral synthesis code LINFOR with model atmospheres calculated with ATLAS9 code \citep[]{ku93}. \citet{beh03} obtained $-1.83 \pm 0.02$ for [Mg/H] and $-2.23 \pm 0.06$ for [Fe/H] relative to the solar value, which are consistent with our results plotted in Figure~$\ref{fig:anomalies}$. The abundance for all the elements found in BD\,+01$^{\circ}$0548 shows an underabundance relative to the solar abundance.

For HD\,74721, we report in Table~$\ref{tab:abhd74721}$ the results for 19 ions. Our results are generally in agreement with those found by \citet{ade96} and \citet{kin00}. Si\,{\sc ii} shows the largest difference ($0.3$ dex) with the value of \citet{ade96}, while the values of \citet{kin00} for Mg\,{\sc ii}, Sc\,{\sc ii}, Ti\,{\sc ii}, Cr\,{\sc ii}, Fe\,{\sc i}, and Fe\,{\sc ii} are consistent with ours within the uncertainties. All elements studied here in the atmosphere of HD\,74721 show an underabundance (see Fig.~$\ref{fig:anomalies}$).

For HD\,86986, Table~$\ref{tab:abhd86986}$ lists the average abundance of 18 ions. In comparison, the average abundance of 14 ions also studied by \citet{ade96} differs from ours by approximately $0.4$ dex. However, Sr\,{\sc ii} shows the largest difference ($1.1$~dex). The average abundance of Mg\,{\sc ii}, Ca\,{\sc i}, Ti\,{\sc ii}, Cr\,{\sc ii}, Fe\,{\sc i}, Fe\,{\sc ii}, and Ba\,{\sc ii} are in agreement with those obtained by \citet{kin00}. All elements show an underabundance (see Fig.~$\ref{fig:anomalies}$).

For HD\,87112, we present in Table~$\ref{tab:abhd87112}$ our results for 14 ions. Our abundance values for Mg\,{\sc ii}, Ti\,{\sc ii}, Cr\,{\sc ii}, Fe\,{\sc i}, and Fe\,{\sc ii} and those of \citet{kin00} are consistent with each other within the uncertainties. \citet{beh03} found an average abundance of $-1.13 \pm 0.06$ for [Mg/H] and $-1.65 \pm 0.07$ for [Fe/H], which are in agreement with ours plotted in Figure~$\ref{fig:anomalies}$. All elements show an underabundance.

For HD\,93329, Table~$\ref{tab:abhd93329}$ gives the average abundance of 23 ions. Except for the neutral elements C\,{\sc i}, N\,{\sc i}, O\,{\sc i}, S\,{\sc i}, and Ni\,{\sc i}, \citet{ade96} analyzed the same ions for this star: Sr\,{\sc ii} shows the greatest difference ($1.5$ dex) followed by  Si\,{\sc ii} and Ca\,{\sc i} ($0.4$ dex). Except for Cr\,{\sc ii}, with a difference of $0.3$ dex, the average abundance found by \citet{kin00} is consistent with ours within the uncertainties. Except for S\,{\sc i}, the abundance of all elements estimated here shows an underabundance (see Fig.~$\ref{fig:anomalies}$).
 
For HD\,109995, we present in Table~$\ref{tab:abhd109995}$ our average abundances for 16 ions. O\,{\sc i} shows the largest difference ($0.9$ dex) followed by N\,{\sc i}, Mg\,{\sc i}, Al\,{\sc i}, Ca\,{\sc i}, Ca\,{\sc ii}, Sc\,{\sc ii}, Cr\,{\sc i}, Fe\,{\sc i}, and Sr\,{\sc ii} (~$0.7$ dex) with the value of \citet{ade87}. The logarithmic value for O\,{\sc i} obtained by \citet{ade86}, and reported in Table~$\ref{tab:abhd109995}$ indicates a clear overabundance ($+0.65$ dex in LTE mode) relative to solar abundance, while their study showed an abundance of $-0.68$ dex for O\,{\sc i} with a non-LTE model. Si\,{\sc ii}, Ti\,{\sc ii}, and Fe\,{\sc ii} show the smallest difference ($0.2$ dex) with \citet{ade87}. The values of  Mg\,{\sc ii}, Ti\,{\sc ii}, Fe\,{\sc i}, and Fe\,{\sc ii} from \citet{kin00} are within our uncertainties.

For HD\,161817, results for 16 ions are shown in Table~$\ref{tab:abhd161817}$. Our values are in agreement with those found by \citet{ade87}, except for Mg\,{\sc ii}, Al\,{\sc i}, N\,{\sc i}, and Ba\,{\sc ii} which display slight differences in their abundances as compared to ours. The logaritmic value of O\,{\sc i} ($-2.62$ ) found by \citet{ade86}, with a LTE model, is in agreement with ours ($-2.39 \pm 0.41$). However, the average abundance of O\,{\sc i} calculated by \citet{ade86} using a non-LTE model is $-3.89$. Six ions (Mg\,{\sc ii}, Ca\,{\sc i}, Ti\,{\sc ii}, Fe\,{\sc i}, Fe\,{\sc ii}, and Ba\,{\sc ii}) were also studied by \citet{kin00} and their results match well with ours.

For HD\,167105 we give in Table~$\ref{tab:abhd167105}$ our average abundance for 11 ions. Five of these ions were studied by \citet{kin00} and are in agreement with our results. The average abundance of seven ions were also calculated by \citet{ade94} and their values generally fit with ours, except for Si\,{\sc ii}, Sc\,{\sc ii}, and Ti\,{\sc ii}. As for HD\,161817, O\,{\sc i} in HD\,167105 shows a clear overabundance (see Fig.~$\ref{fig:anomalies}$).

\subsection{Vertical abundance stratification}

To evaluate the presence of vertical abundance stratification for a given element, we consider that at least 10 lines must be visible over a broad range of optical depths (approximately two orders of magnitude for $\tau_{5000}$).  Among the elements selected in the atmosphere of our eight field BHB stars, iron shows the largest number of unblended lines, followed by titanium. Tables~$\ref{tab:abbd010548}$ to $\ref{tab:abhd167105}$ give for each star the number of lines ($N$) studied for each ion identified in their spectrum.  To declare an element as stratified, we considered that: 1) the slope of the variation of the abundance as a function of the optical depth must be statistically significant, that is greater than 3$\sigma$; and 2) the overall abundance variation must be larger than $0.5$ dex. The second criterion ensures that the abundance varies more than what might be expected due to uncertainty, which can be caused by various factors. Taking these criteria into account, no clear vertical stratification is detected.

\begin{figure*}[ht!]
\centering
\includegraphics[scale=0.70]{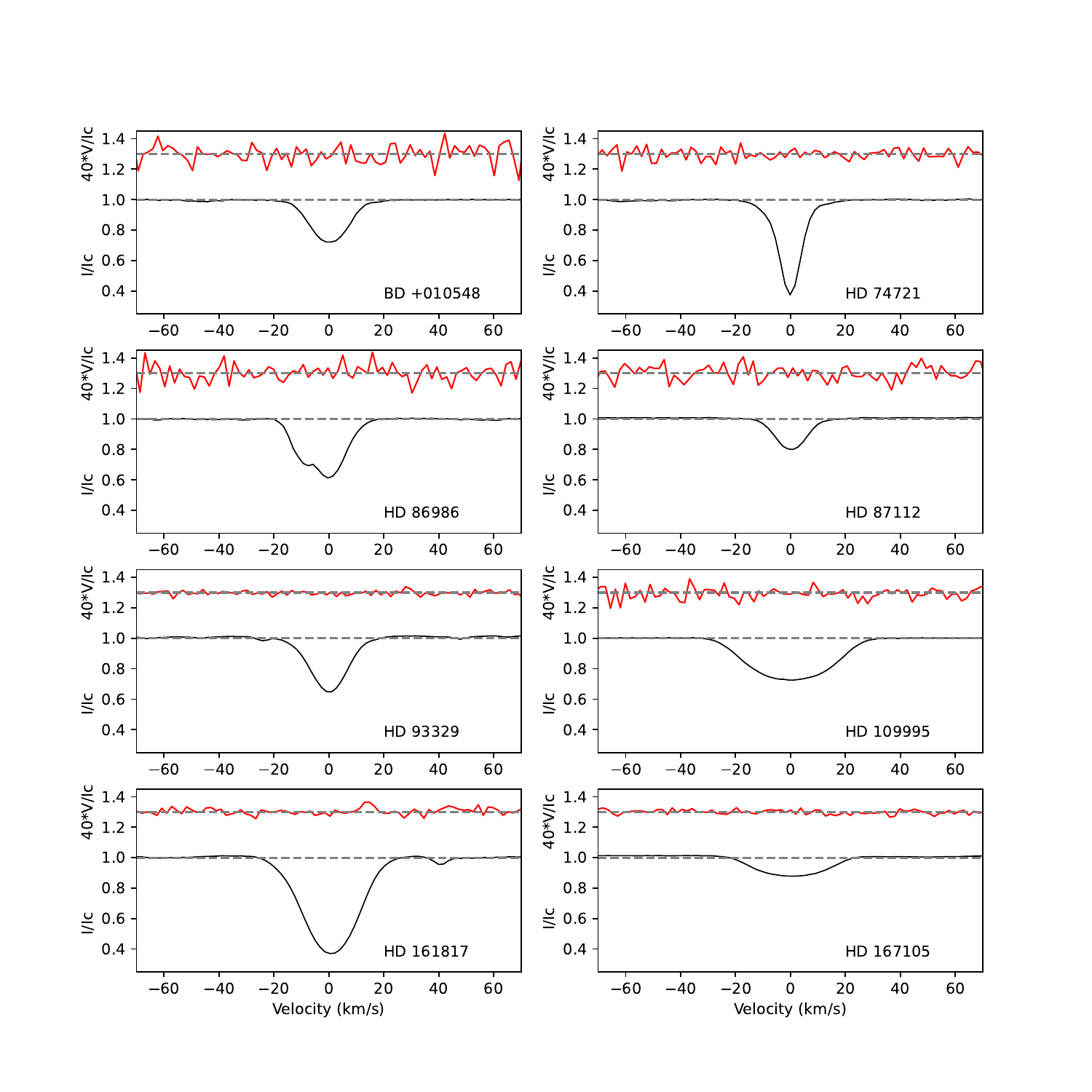}
\caption{Illustration of the LSD unpolarized (Stokes $\mathrm{I}$; bottom curve) and circularly polarized (Stokes $\mathrm{V}$; top curve) profiles for the stars studied. The circular polarization signature is magnified by a factor of 40 times.}
\label{fig:profilbhb}
\end{figure*}

\begin{deluxetable*}{lccc}
%\digitalasset
\tablewidth{700pt} 
\tablecaption{Longitudinal magnetic field measurements obtained and measured detection probability.\label{tab:magnetic}}
\tablehead{
\colhead{Star} & \colhead{This study} &\colhead{This study} & \colhead{Elkin (1998) } \\
 & \colhead{$B_{\ell} \pm \sigma$ (G)}& \colhead{Detect Prob $(\%)$} & \colhead{$B_{\ell} \pm \sigma$ (G)}
}
\startdata 
BD\,+01$^{\circ}$0548  &  $+9 \pm 30$  &  $ 1.07$ & \nodata \\
HD\,74721 &   $-8 \pm 14$  &  $0.05 $ & $+240 \pm 150$ \\
HD\,86986 &   $-15 \pm 23$  &  $1.07 $ & $-430 \pm 580$  \\
HD\,87112 &   $-12 \pm 26$  &  $3.13 $ & \nodata \\
HD\,93329 &   $4 \pm 8$  &  $0.36 $ & \nodata \\
\hline
HD\,109995 &   $-13 \pm 16$  &  $2.14 $ & $-820 \pm 470$ \\
{} &   \nodata  &  \nodata & $+700 \pm 520$ \\
\hline
HD\,161817  &   $-5 \pm 10$  &  $2.16 $ & $+30 \pm 110$\\
{}  &   \nodata  &  \nodata & $-550 \pm 140$\\
{}  &   \nodata  &  \nodata & $-90 \pm 80$\\
{}  &   \nodata  &  \nodata & $+100 \pm 160$\\
\hline
HD\,167105 &   $+23 \pm 19$  &  $3.38 $ & \nodata \\          
\enddata
\end{deluxetable*}

\section{MAGNETIC FIELD MEASUREMENTS}

To search for magnetic fields in the atmosphere of our targets, we used the Least-Squares Deconvolution (LSD) technique which performs a cross-correlation of order to detect and measure very small polarization signatures which indicate the presence of a magnetic field \citep[]{don97}. 

We obtained, from the database VALD ({\tt extract stellar requests}), line masks for each star (considering their physical parameters as given in Tab.~\ref{tab:parameters}). From the VALD linelists, we removed the Balmer lines, all lines blended with the Balmer lines, and those blended with telluric lines. The last step of this exercise consisted of adjusting the mask depths to properly fit the mean observed line depths. This process produced for each observed spectrum mean Stokes $\mathrm{I}$ and $\mathrm{V}$ profiles. 

We used the iLSD code of \citet{2010A&A...524A...5K} to apply the LSD technique to all our observed spectra. Using the adjusted line masks (the final mask that results in the best fit between the LSD model and the observed Stokes I spectrum), the iLSD code allows for the extraction of a mean Stoke $\mathrm{I}$, a mean Stoke $\mathrm{V}$ and null profiles on a velocity grid with a resolution of 1.8~km~s$^{-1}$. Figure~$\ref{fig:profilbhb}$ shows the mean LSD Stokes $\mathrm{V}$ and Stokes $\mathrm{I}$ profiles computed for the stars studied.

The mean longitudinal magnetic field ($B_{l}$) is the line-of-sight component of the stellar magnetic field, suitably weighted and integrated over the visible stellar disc. It can be represented as the first-order moment of the Stokes $\mathrm{V}$ profile \citep[]{don97,wa00}. To determine the mean longitudinal magnetic field ($B_{l}$ in gauss) of each star, we used the following equation:
\begin{equation}
B_{\ell} = 2.14\times10^{11} \frac{\int v \mathrm{V}(v) dv}{\lambda g c {\int\{1 - \mathrm{I}(v)\}dv}},
\end{equation}
where $c$ is the speed of light, $g$ is the average Land\'{e} factor and $\lambda$ (in nm) is the wavelength of the LSD profile, $\mathit{v}$ is the radial velocity. $\mathrm{V}(\mathit{v})$ and $\mathrm{I}(\mathit{v})$ are, respectively, the Stokes $\mathrm{V}$ and Stokes $\mathrm{I}$ parameters \citep[]{don97}. The uncertainty obtained with each measurement is calculated from the statistical error bars of the photons propagated through the reduction of the polarization spectra and the computation of the LSD profiles \citep[]{wa00}.

Table~$\ref{tab:magnetic}$ presents our results of the longitudinal magnetic field ($B_{\ell}$) for the stars studied, as well as the probability of detection of the magnetic field.  None of our results provide any suggestion of a magnetic field. 

We also report in Table~\ref{tab:magnetic} the values of $B_{\ell}$ measured by \citet{Elk98} for four stars. His results are consistent with ours with the non-detection of a magnetic field. \citet{Elk98} used Zeeman spectra from the 6-metre telescope of the Special Astrophysical Observatory of the Russian Academy of Sciences and obtained values of the longitudinal magnetic field with uncertainties ranging from 8 to 32 times ours.

\section{DISCUSSION AND CONCLUSION}

\subsection{Chemical abundances and stratification}

We analyze the spectrum of eight field  BHB stars (BD\,+01$^{\circ}$0548, HD\,74721, HD\,86986, HD\,87112, HD\,93329, HD\,109995, HD\,161817, and HD\,167105) observed with the echelle spectropolarimeter ESPaDOnS at the CFHT. This spectral analysis project focused on a study of average abundances, vertical stratification of chemical abundances, and of a search for magnetic fields in the atmosphere of the stars.

Firstly, we used the ZEEMAN2 code and PHOENIX model atmospheres to determine the average abundance of chemical elements, the radial velocity, and the rotational velocity. Generally, our results of the average element abundances are consistent with those found in previously published studies. No clear signatures of vertical stratification of iron or titanium were detected here in the atmosphere of these stars. An overabundance of O\,{\sc i} was detected in HD\,161817 and HD\,167105. All other elements analyzed are underabundant. 

\subsection{Magnetic fields}

As described above, BHB stars are descendants of low-mass (0.5 - 1~M$_\odot$) main-sequence stars that have ascended the red-giant branch, undergone a He flash, and settled into He core fusion on the horizontal-branch. Their temperatures corresponding to spectral types A and B imply that their atmosphere is principally radiative. In the context of ongoing discussions of the origin of the strong ``fossil” magnetic field detected in some main-sequence A and B stars \citep[e.g.][]{1980ApJS...42..421B} and post-main-sequence stars \citep[e.g.][]{2023Sci...381..761S}, these fields are characterized by relatively simple surface topologies with important dipolar components that are rather strong (generally more than a few hundred gauss, and frequently several kilogauss) and stable on at least observational timescales (i.e. decades). 

Models suggest that a process of general relaxation of stochastic initial fields can lead to certain stable relaxed field configurations of mixed toroidal and poloidal components \citep[e.g.][]{2004Natur.431..819B}. The origin of these seed fields has been the subject of enthusiastic debate and investigation. Among suggested sources are advection of interstellar magnetic field during star formation, remnant field from pre-main-sequence dynamos, and even binary interaction via mass transfer and mergers. 

Given that their red-giant progenitors are known to have supported strong dynamo-driven magnetic fields \citep[e.g.][]{2015A&A...574A..90A,2025BlgAJ..42...35K}, BHB stars appear to be a natural testbed for the suggestion that progenitor dynamo fields may serve as an origin of fossil fields in their radiative descendants. Previously, \cite{Elk98} performed an analogous polarimetric analysis for magnetic fields in BHB stars, obtaining longitudinal field measurements of eight field horizontal-branch stars with uncertainties typically larger than 100~G. For one star, HD~161817, his results seem to indicate the possible presence of a stable magnetic field.

Other investigations \citep[e.g.][]{2019A&A...622A..77P} sought indirect evidence for magnetic fields without any positive results. 

In this project we have obtained single circular polarization (Stokes $\mathrm{V}$) sensitive to longitudinal magnetic fields of a few 10s of gauss. Our observations rule out longitudinal magnetic fields stronger than 100~G in all cases with high confidence (4 - 10$\sigma$). Our sample includes several targets previously observed by \citet{Elk98}. Our re-observation of HD~161817 yields a null result with an error bar of 10~G. Although re-observation would be beneficial, this null result suggests that the previously reported detection with much larger uncertainties (80-160~G) may be spurious.

Given the lack of detection in any of our targets, we are unable to make any strong general comments on the potential generation of fossil fields due to progenitor-dynamo activity. Our immediate take-away is that strong magnetic fields with longitudinal field larger than a few 10s of gauss are rare in BHB stars. Nevertheless, such fields are also rare in main-sequence stars, being detected in only about 1 in 10 objects \citep[e.g.][]{1968PASP...80..281W}. The small size of our sample is an obvious limitation, and future observations should seek to expand the sample to allow stronger conclusions to be drawn.

%\section*{ACKNOWLEDGMENTS}
\begin{acknowledgments}
The authors thank Digital Research Alliance of Canada for computational resources and CFHT for allocation of observing time for this project. This work has used the VALD database, operated at Uppsala University, the Institute of Astronomy RAS in Moscow, and the University of Vienna. Carmelle Robert is grateful to NSERC for financial support.
\end{acknowledgments}

\bibliography{FBHB}{}
\bibliographystyle{aasjournal}
%\textbf{SUPPORTING INFORMATION}\\
%Additional Supporting Information may be found in the online version of this paper:\\
%\textbf{Table 3a} - BD\,+010548\\
%\textbf{Table 3b} - HD\,74721\\
%\textbf{Table 3c} - HD\,86986\\
%\textbf{Table 3d} - HD\,87112\\
%\textbf{Table 3e} - HD\,93329\\
%\textbf{Table 3f} - HD\,109995\\
%\textbf{Table 3g} - HD\,161817\\
%\textbf{Table 3h} - HD\,167105

\end{document}